\documentclass[pdflatex,sn-mathphys]{sn-jnl}

\jyear{2021}%

\usepackage{amsmath}
\usepackage{amsthm}
\usepackage{natbib}
\theoremstyle{plain}

\usepackage{mathtools}

\newtheorem{thm}{Theorem}[subsection]
\newtheorem{prop}[thm]{Proposition}
\newtheorem{lem}[thm]{Lemma}
\newtheorem{cor}[thm]{Corollary}
\theoremstyle{definition}
\newtheorem{dfn}[thm]{Definition}

\newcommand{\bprop}[1]{ \begin{prop} #1 \end{prop}}

\newcommand{\bdfn}[1]{ \begin{dfn} #1 \end{dfn}}
\newcommand{\bproof}[1]{ \begin{proof} #1 \end{proof}}

\usepackage{amssymb}
\usepackage{listings}
\usepackage{comment}
\usepackage{enumerate}
\usepackage{physics}

\newcommand{\argmin}{\mathop{\rm arg~min}\limits}

\newcommand{\eq}[1]{\begin{align*} #1  \stepcounter{equation}\tag{\theequation} \end{align*}}
\newcommand{\SM}[1]{\overline{\overline{{#1}}}}
\newcommand{\opnorm}[1]{ \norm{#1}_{\rm op}}

\def \({\left(}
\def \){\right)}
\def \[{\left[}
\def \]{\right]}
\def \.{~.}
\def \,{~,}

\newcommand{\NP}{ \textsf{NP}}

\newcommand{\NPh}{ \textsf{NP}\textrm{-hard}}
\newcommand{\BQP}{ \textsf{BQP}}

\newcommand{\hH}{\hat{H}}
\newcommand{\hU}{\hat{U}}
\newcommand{\hM}{\hat{M}_x}
\newcommand{\hr}{\hat{\rho}}
\newcommand{\ga}{\gamma}
\newcommand{\gb}{(\gamma)}
\newcommand{\gba}{(\gamma;\alpha)}
\newcommand{\al}{\alpha}
\newcommand{\ep}{\epsilon}
\newcommand{\gbi}{\tilde{\gamma}}
\newcommand{\mbi}{\tilde{m}_x^\NJ}
\newcommand{\alN}{\frac{\al}{N}}

\newcommand{\NJ}{{N:J}}
\newcommand{\ABJ}{{A\# B:J}}
\newcommand{\AkJ}{{A[k]:J}}

\newcommand{\AoneJ}{{A[1]:J}}
\newcommand{\kdB}{ (k = 1, \dots ,B)}
\newcommand{\kdBA}{ (k = 1, \dots ,B = N/A)}
\newcommand{\AD}{ A^{1-1/D}}
\newcommand{\Mx}{ (M_x)}
\newcommand{\mx}{ (Nm_x)}
\newcommand{\Mxa}{ (M_x;\al)}
\newcommand{\mxa}{ (Nm_x;\al)}
\newcommand{\cbulk}{c_{\mathrm{bulk}}}
\newcommand{\csurface}{c_{\mathrm{surface}}}

\newcommand{\Mh}{ \hM^N}
\newcommand{\MhAk}{ \hat{M}_x^{A[k]}}
\newcommand{\MhAl}{ \hat{M}_x^{A[l]}}

\newcommand{\Hg}{ E_g^\NJ}
\newcommand{\Hh}{ \hH^\NJ}
\newcommand{\Ug}{ U_g^\NJ}
\newcommand{\Uh}{ \hU^\NJ}

\newcommand{\HgAB}{ E_g^\ABJ}
\newcommand{\HhAB}{ \hH^\ABJ}
\newcommand{\UgAB}{ U_g^\ABJ}
\newcommand{\UhAB}{ \hU^\ABJ}
\newcommand{\UhAk}{ \hU^\AkJ}

\newcommand{\Hgg}{ \Hg \gb}
\newcommand{\Hhg}{ \Hh \gb}
\newcommand{\hgg}{ e_g \gb}
\newcommand{\UgM}{ \Ug \Mx}
\newcommand{\Ugm}{ \Ug \mx}
\newcommand{\ugm}{ u_g(m_x)}

\newcommand{\Uha}{ \hU^\NJ(;\al)}

\newcommand{\Hhga}{ \Hh \gba}

\newcommand{\UgMa}{ \Ug \Mxa}
\newcommand{\Ugma}{ \Ug \mxa}
\newcommand{\ugma}{ u_g(m_x;\al)}

\newcommand{\sumi}{ \sum_{i=1}^N}
\newcommand{\sumj}{ \sum_{j=1}^N}
\newcommand{\sumk}{ \sum_{k=1}^B}
\newcommand{\suml}{ \sum_{l=1}^B}

\newcommand{\minr}{ \min_{\hr}}
\newcommand{\minM}{ \min_{\hr \mid \Tr \hr \Mh = M_x}}

\newcommand{\argminr}{ \argmin_{\hr}}

\newcommand{\argminm}{ \argmin_{\hr \mid \Tr \hr \Mh = Nm_x}}
\newcommand{\maxml}{ \max_{m_x \in [0 , \mbi]}}

\newcommand{\maxgl}{ \max_{\ga \in [0 , \gbi]}}

\newcommand{\supll}{\sup_{\substack{\ga, \ket{g} \ {\rm s.t.} \\  Nm_* \in I_1 , M_*^\NJ \in I_1}}}
\newcommand{\suplr}{\sup_{\substack{\ga, \ket{g} \ {\rm s.t.} \\ Nm_* \in I_1 , M_*^\NJ \in I_2}}}
\newcommand{\suprl}{\sup_{\substack{\ga, \ket{g} \ {\rm s.t.} \\ Nm_* \in I_2 , M_*^\NJ \in I_1}}}
\newcommand{\suprr}{\sup_{\substack{\ga, \ket{g} \ {\rm s.t.} \\ Nm_* \in I_2 , M_*^\NJ \in I_2}}}

\begin{document}

\title[Proof of avoidability of the QFOT in QA of finite-dimensional spin glasses]{Proof of avoidability of the quantum first-order transition in transverse magnetization in quantum annealing of finite-dimensional spin glasses}


\author*[1]{ \sur{Mizuki} \fnm{Yamaguchi}}\email{yamaguchi-q@g.ecc.u-tokyo.ac.jp}

\author[1]{\sur{Naoto} \fnm{Shiraishi} }

\author[1]{\sur{Koji} \fnm{Hukushima} }

\affil[1]{\orgdiv{Graduate School of Arts and Sciences}, \orgname{The University of Tokyo}, \orgaddress{\street{Meguro-ku 3-8-1}, \city{Tokyo}, \postcode{1530041},  \country{Japan}}}





\abstract{It is rigorously shown that an appropriate quantum annealing for any finite-dimensional spin system has no quantum first-order transition in transverse magnetization. 
This result can be applied to finite-dimensional spin-glass systems, where the ground state search problem is known to be hard to solve.
Consequently, it is strongly suggested that the quantum first-order transition in transverse magnetization is not fatal to the difficulty of combinatorial optimization problems in quantum annealing.}




\maketitle
\section{Introduction}
Solving combinatorial optimization problems efficiently is a major topic in theoretical computer science. 
From the viewpoint of physics, this problem can be described as energy minimization with a given classical spin system. 
Inspired by this, the simulated annealing~\cite{kirkpatrick1983optimization} and the quantum annealing (QA)~\cite{kadowaki1998quantum, farhi2000quantum} were invented as generic heuristic methods for solving these optimization problems. 
Given a classical Hamiltonian $\Uh$ of system size $N$ and for the set of quenched coupling constants $J$, the QA solves the energy minimization problem as follows:
we set the Hamiltonian of the quantum system as
\eq{ \Hhg = \Uh + \ga \hat{D} \, }
where $\hat{D}$ is noncommutative with $\Uh$ and $\gamma$ is a parameter controlled in QA. 
This Hamiltonian $\hat{D}$ is called the {\it driver Hamiltonian} and has a known ground state. 
Starting with the system with large $\ga$ and varying slowly to $\ga=0$, one would expect the ground state of $\Hhg$ to change from the known ground state of $\hat{D}$ to the desired solution state, the ground state of $\Uh$.
The quantum adiabatic theorem~\cite{kato1950adiabatic} guarantees that the desired ground state is attainable by taking a sufficiently long time to change $\ga$. 
A natural question raised here is whether this process is time efficient, i.e., whether it takes only polynomial time of the system size $N$ to reach the true ground state of $\Uh$. 
Roughly speaking, the time cost is proportional to the inverse square of the minimum gap between the ground-state energy and the first excited state energy of the Hamiltonian in the annealing process~\cite{farhi2000quantum, albash2018adiabatic}.

The first proposal of QA employed the transverse magnetic term $- \Mh = - \sumi \hat{\sigma}_i^x$ as the driver Hamiltonian $\hat{D}$, and argued based on numerical experiments that the QA with this driver Hamiltonian is superior to simulated annealing~\cite{kadowaki1998quantum}. 
On the other hand, this type of QA was shown to fail in the $p$-spin model ($p$-body mean-field ferromagnetic model with $p>2$)~\cite{jorg2010energy,seki2012quantum}, where the gap is exponentially small and the computation time is exponentially large. 
In the case of the $p$-spin model, the transverse magnetization undergoes a discontinuous jump in the annealing process, which we call the quantum first-order transition in transverse magnetization (QFOT for short) analogous to the first-order transition in thermodynamics.
This is a challenging phenomenon for QA, since the sudden change in the transverse magnetization causes an exponential collapse of the gap, implying the failure of annealing.

The fundamental origin of the failure of the QA for hard optimization problems is under discussion in this field.
One possible argument is that the failures mainly are due to the QFOT~\cite{jorg2010first}, based on the observed fact that the QFOT frequently appears when the QA fails~\cite{seki2015quantum}.
In contrast, another argument is that the failure of the QA has origins other than the QFOT.
For example, it was reported in Ref~\cite{young2008size,young2010first,farhi2012performance,knysh2016zero,takahashi2019phase} that QA in some models shows exponentially small gaps at points other than the point of QFOT. 
However, the latter observation is based on specific models and does not provide a general argument on the relation between the QFOT and the failure of the QA.


To resolve this controversy, we adopt the QA with antiferromagnetic fluctuations (QA-AFF) first proposed in Ref~\cite{seki2012quantum} and set the entire Hamiltonian as $\Hhga = \Uh - \ga \Mh + \alN (\Mh)^2$.
The additional fluctuations term makes the QFOT more avoidable.
Previous studies based on specific models show that in the ferromagnetic $p$-spin model (without quenched randomness) and the Hopfield model, the QA-AFF succeeds in avoiding QFOT under some conditions, while the QFOT appears unavoidable under others~\cite{seki2012quantum, seki2015quantum}.
In spite of these previous investigations, the potential effectiveness of the QA-AFF in general systems has not yet been uncovered.

In this paper, we rigorously prove that the QA-AFF for finite-dimensional spin-glass systems avoids the QFOT. 
Our result applies to general finite-dimensional spin-glass systems as long as i.i.d. quenched random variables are used.
Even for systems that exhibit the QFOT under the conventional QA with only the transverse magnetic field, the addition of the antiferromagnetic fluctuations term always removes this singularity and makes the transverse magnetization in this QA continuous as a function of $\ga$.
In other words, the QFOT in QA can be completely avoided by adding antiferromagnetic fluctuations.

We note that the search for the ground state of three-dimensional 
Ising spin-glass systems is considered to be a computationally hard task since it belongs to an $\NPh$ problem~\cite{barahona1982computational}, which is a class of the most difficult combinatorial optimization problems. 
It is believed that even quantum computers cannot solve $\NPh$ problems efficiently, which suggests that the QA-AFF in fact fails at some point.
Based on our findings, we assert that the QFOT is not fatal to the difficulty of combinatorial optimization problems in QA.

In our proof, the self-averaging plays a pivotal role to derive the absence of singularity.
First, we show that the ground-state energy $\Hgg$ is self-averaging in the conventional QA only with the transverse magnetic field.
There remains the possibility that the function obtained by taking the average of the quenched randomness has a singularity (i.e., non-differentiable points).
Then, it can be shown that the addition of the antiferromagnetic fluctuations term does indeed remove the singularity for any finite-dimensional system. However, for some long-range interaction systems, this claim may not hold, which is consistent with the previous studies showing that QFOT cannot be avoided in the mean-field models.

This paper is organized as follows. In Section~\ref{setupclaim}, we explain our setup and main claim. 
In Section~\ref{outline}, we describe the outline of the proof, presenting several key ideas in this proof. 
In Section~\ref{proof}, which consists of four subsections, we prove the main theorem. 
We briefly review the Legendre transformation in Subsection~\ref{preparation}.
Subsection~\ref{Self-averaging} and \ref{uniform-self-averaging} are devoted to the investigation of self-averaging for a fixed parameter and uniform self-averaging for a function, respectively.
We finally introduce the antiferromagnetic fluctuations in Subsection~\ref{subaff}, which completes the proof of the avoidance of the QFOT.

\section{Setup and main claim}
\label{setupclaim}

\subsection{Setup}

We deal with the energy minimization problem of a classical spin-$1/2$ system on a finite-dimensional lattice with $N$ spins. 
Pairs of spins $\hat{\sigma}_i^z$ and $\hat{\sigma}_j^z$ interact with each other through the coupling constant $J_{ij}$, which is a quenched random variable. 
The Hamiltonian of this classical system is thus expressed as
\eq{
\Uh = - \sumi \sumj J_{ij} \hat{\sigma}_i^z \hat{\sigma}_j^z \,
}
where the set of coupling constants $J_{ij}$ is denoted simply by $J$.

To solve the energy minimization problem by QA, we add a transverse magnetic field with strength $\gamma\in (-\infty,\infty)$, which leads to the following Hamiltonian for the quantum system:
\eq{
\Hhg = \Uh - \ga \Mh = \Uh - \ga \sumi \hat{\sigma}_i^x \.
}
As discussed in detail in Sec.~\ref{subaff}, dealing with the QA-AFF, $\Uh$ in this formula is replaced by $\Uh + \frac{\al}{N} (\Mh)^2$.

Throughout this paper, we consider models in finite dimension with $\Uh$ generated by a shift-invariant probability distribution for quenched random variables.
We here clarify the meaning of {\it finite dimension} for later use.
If the lattice is placed in $\mathbb{R}^n$ space and each site interacts only with its neighbors, the notion of dimension has no confusion.
The problem may arise when distant sites also interact with forces that decay with distance.
To cover these systems, we define the finite dimensionality for $D\geq 2$ with $D$ being the spatial dimension as follows\footnote{
In one dimension, the same argument holds but the order evaluation of the result differs.
}:

\bdfn{[Finite dimensionality]
We say that a Hamiltonian $\Uh$ (or a system) is in a finite dimension if the following two conditions are satisfied\footnote{
Our results hold for finite-dimensional systems with general $p$-body interactions. For example, in three-body interacting system $\Uh = - \sum_{h=1}^N \sumi \sumj J_{hij} \hat{\sigma}_h^z \hat{\sigma}_i^z \hat{\sigma}_j^z$, finite-dimensionality is the condition that $\sumi \sumj \sqrt{[ J_{hij}^2 ]} \leq \cbulk$ instead of \eqref{bulk-cond} and that $ \sum_{h \in s(A[k])} \sum_{i \notin s(A[k])} \sum_{j \notin s(A[k])}\sqrt{[ J_{hij}^2 ]} \leq \csurface \AD$ instead of \eqref{surface-cond}.}:
\begin{itemize}
\item For any site $i$, the sum of interactions with $i$ is bounded from above as 
\eq{\label{bulk-cond}
\sumj \sqrt{\left[J_{ij}^2\right]} \leq \cbulk \, 
}
where $[\cdot ]$ is the random average with respect to the quenched randomness (see Subsection~\ref{Self-averaging} for details), and $\cbulk$ is a constant independent of the system size $N$.

\item For any size $A$ of the system, there exists a decomposition of sites into $B = N/A$ subsystems of size $A$ denoted by $A[1], \ldots, A[B]$, such that the sum of interactions between inside and outside of any subsystem $A[k]$ is bounded from above as
\eq{\label{surface-cond}
\sum_{i \in s(A[k])} \sum_{j \notin s(A[k])}\sqrt{\left[J_{ij}^2\right]}  \leq \csurface A^{1-1/D} \,
}
which means that the surface energy is insignificant compared to the bulk energy for a sufficiently large system.
Here $s(A[k])$ is the set of the sites in $A[k]$, and $\csurface$ is a constant independent of the size of the system and the subsystems. 
\end{itemize}
}

We also clarify the meaning of {\it shift-invariant probability distribution} for quenched random variables.
Let $P_{ij}(J_{ij})$ be the probability distribution for the coupling constants $J_{ij}$.
This probability distribution is shift-invariant if $P_{ij}=P_{kl}$ holds for any $i,j,k,l$ such that $\mathbf{r}_i-\mathbf{r}_j=\mathbf{r}_k-\mathbf{r}_l$, where $\mathbf{r}_i$ is a $D$ dimensional vector representing the lattice position of site $i$. 
If the lattice consists of several sublattices, the above characterization further requires that $i$ and $k$ belong to the same sublattice. The system without quenched randomness is considered a special case of a shift-invariant system.

\subsection{Main result}
We shall define the quantum first-order transition in transverse magnetization (QFOT) in systems with quenched random variables. 
We first provide the definition of the absence of the QFOT for a system without quenched randomness, i.e., $\Uh$ is deterministically constructed depending on $N$.
In this case, we say that the QA does not exhibit a QFOT if the transverse magnetization density $\frac{\ev{\Mh}{g(\ga)}}{N}$ of a ground state $\ket{g(\ga)}$ of $\Hhg$ converges uniformly to a continuous function $m_*(\ga)$ in thermodynamic limit ($N \to \infty$):
\eq{
\lim_{N \to \infty}  \sup_{\ga \in [0,\infty)} \abs{ \frac{\ev{ \Mh }{g(\ga)}}{N} - m_*(\ga) } = 0 \.
}
Here, the ground state of $\Hhg$ is implicitly assumed to be unique.
If the ground state degenerates, we regard that our condition should be satisfied for any state $\ket{g}$ in the state space of minimum energy: $G(\Hhg) = \argmin_{\ket{\psi}} \ev{\Hhg}{\psi}$.
Or equivalently, the above definition can be replaced by
\eq{
\lim_{N \to \infty} \sup_{\ga \in [0,\infty)} \max_{ \ket{g} \in G(\Hhg)} \abs{ \frac{\ev{ \Mh }{g}}{N} - m_*(\ga) } = 0 \.
}

In the case of spin glasses with quenched randomness, we shall define the QFOT as its typical behavior, replacing the convergence in the above definition with stochastic convergence (especially convergence in mean square).

\bdfn{
We say that a given QA shows no quantum first-order phase transition in transverse magnetization if there exists a continuous function $m_*(\ga)$ such that
\eq{
\lim_{N \to \infty} \[ \sup_{\ga \in [0,\infty)} \(\frac{\ev{ \Mh }{g(\ga)}}{N} - m_*(\ga) \) ^2 \] = 0 \.
}
Here, $[\cdot ]$ means the average for the quenched random variable.
}

We remark that the function $m_*(\ga)$ is independent of the quenched random variables.
Thus, the above definition states that for almost all $\Uh$ the transverse magnetization density in the ground state converges to the same function.

We prove that the QFOT in the above definition is always avoidable in finite-dimensional spin-glass systems.
\begin{thm}
\label{main_theorem}
For any classical Hamiltonian in finite dimension with quenched random variables sampled from a shift-invariant probability distribution, the QA-AFF for this Hamiltonian does not have a QFOT.
\end{thm}
This is the main result of this paper.
Notice that this theorem only describes the absence of a jump in the transverse magnetization density during the QA process, and does not evaluate the size of the energy gap that determines the annealing time required for successful computation.

\subsection{Remark}

We here put two remarks related to our main result.

\medskip

The first remark is on the connection to the computational hardness, in particular computational complexity.
It is known that the energy minimization problem in three-dimensional spin glass is an $\NPh$ problem. 
Thus, from the perspective of computational complexity, the finite-dimensional spin glass with $D\geq 3$ is as complex as the Sherrington-Kirkpatrick model.
Therefore, our subject is indeed the efficiency of QA for hard combinatorial optimization problems.

This paper rules out the possibility that the difficulty in QA for finite-dimensional spin glasses lies in the QFOT. 
Our result, however, does not state that the QA-AFF actually succeeds as a method for solving combinatorial optimization problems and that $\NP \subseteq \BQP$ in terms of computational complexity. 
A more plausible scenario would be that the QA-AFF suffers from causes other than the QFOT.
We will discuss this point in detail in Section~\ref{discussion}.

\medskip

The second remark is on the form of $\Uha$ in QA-AFF.
In QA-AFF, the Hamiltonian at $\ga = 0$ is $\Uha=\Uh + \frac{\al}{N} (\Mh)^2$, not the classical spin glass Hamiltonian $\Uh$ we wish to solve. 
However, this discrepancy does not matter for the success or the failure of the QA for the following reason:
if the coupling constants do not take continuous real numbers but discrete numbers of decimal places, by setting $\al$ smaller than the smallest unit of energy and measuring the final state with the computational basis, we can observe one of the true ground states with a finite probability\footnote{
Let $\ket{{\rm ans}}$ be a classical ground state of $\Uh$ and $d$ be the smallest unit of energy of $\Uh$. Since we have $\ev{\Uh + \alN (\Mh)^2}{{\rm ans}} = \ev{\Uh}{{\rm ans}} + \al$, the ground-state energy of $\Uh + \alN (\Mh)^2$ is lower than that of $\Uh$ plus $\al$. Consequently, measuring the ground state of $\Uh + \alN (\Mh)^2$ with the computational basis yields the ground state(s) of $\Uh$ with probability at least $1 - \al / d$ if $\al < d$.
}. 
Indeed, under the restriction that the coupling constants are integers, ground-state search problem for three-dimensional spin glasses has been proven to be $\NPh$~\cite{barahona1982computational}.

\section{Outline of the proof}
\label{outline}
Our first simple but important step is inspired by thermodynamics, where thermodynamic functions (e.g., the Helmholtz free energy and the Gibbs free energy) are connected via the Legendre transformation with respect to an extensive variable (e.g., volume) and an intensive variable (e.g., pressure).
We regard the ground-state energy $\Hgg$ as a \textit{thermodynamic function} with an \textit{intensive variable} $\ga$.
Then, its inverse Legendre transformation~\cite{shimizu2021, callen1991thermodynamics} yields the constrained minimum energy $\UgM$ with an \textit{extensive variable} $M_x$.

The functional forms of $\Hgg$ and $\UgM$ depend on $\Uh$. 
However, if the probability distribution generating $\Uh$ is shift-invariant, we can show that
\eq{
\frac{\Hgg}{N} &\sim \hgg \, \\
\frac{\Ugm}{N} &\sim \ugm \, 
}
where $\sim$ stands for stochastic convergence (especially convergence in mean square), which is nothing but self-averaging in terms of physics.
In the conventional QA only with a transverse magnetic field, $-\hgg$ and $\ugm$ are shown to be convex, while $\ugm$ might be not strictly convex, which leads to a singularity as an non-differentiable point in $\hgg$ at the QFOT.

To remove this singularity, we add the antiferromagnetic fluctuations term to the Hamiltonian $\Hhg$, which results in $\Hhga = \Uh - \ga \Mh + \frac{\al}{N} ( \Mh )^2$. 
In this case, the corresponding $\ugma$ is strictly convex, since $\ugm$ is convex and the square function $\al m_x^2$ provides a small convex curve. 
Consequently, it follows that the QFOT does not occur in a typical evaluation with the addition of the antiferromagnetic fluctuations term.

The effort for this proof is mainly devoted to proving self-averaging. In this paper, we use the finite dimensionality for the proof, while some studies employ other methods~\cite{chatterjee2015absence}. To show the self-averaging of $\Hgg$, we decompose the system into $B = N/A$ subsystems of size $A$ and use a version of the law of large numbers recalling that these subsystems are i.i.d. Subsequently, we show the uniform self-averaging (uniform convergence) of $\Hgg$. The assertion of the main theorem~\ref{main_theorem} is described as uniform self-averaging of the transverse magnetization.

\section{Proof}
\label{proof}
\subsection{Preparation}
\label{preparation}

We introduce two functions $\Hgg$ and $\UgM$ analogous to \textit{thermodynamic functions} and demonstrate that they are Legendre transformation and inverse Legendre transformation~\cite{shimizu2021} of each other. 
For completeness, below we will describe several basic results of the Legendre transformation in terms of $\Hgg$ and $\UgM$.
Readers who are familiar with these techniques can skip this subsection.

\bdfn{The ground-state energy of $\Hhg$ is defined as
\eq{
\Hgg &= \min_{\ket{\psi}} \ev{\Hh \gb}{\psi} \.
}
Here, $\ket{\psi}$ runs all possible pure states.
}

The domain of minimization in the above definition can be extended from pure states to general mixed states.

\bprop{The ground-state energy $\Hgg$ is also the minimum expectation energy of general mixed states:
\eq{
\Hgg = \minr \Tr \hr \Hhg \.
}
}
\begin{proof}
Fix $\hr' \in \argminr \Tr \hr \Hhg$.
Decomposing this state as $\hr' = \sum_t p_t \ketbra{\psi_t}$, we obtain
\eq{
\minr \Tr \hr \Hhg
&= \Tr \hr' \Hhg \\
&= \sum_t p_t \ev{\Hhg}{\psi_t} \\
&\geq \Hgg \.
}
The inverse inequality is obvious.
\end{proof}

We next introduce the minimum energy conditioned by the transverse magnetization.

\bdfn{The minimum energy of $\Uh$ conditioned by $x$-magnetization at $M_x=\ev{\Mh}$ is defined as
\eq{
\UgM &= \minM \Tr \hr \Uh \.
}
}

Notably, $\Hgg$ and $\UgM$ are connected through the Legendre transformation.

\bprop{
\label{legendre1}
$\Hgg$ is the Legendre transformation of $\UgM$.
}
\bproof{Combining the definitions of $\Hgg$ and $\UgM$, we find
\eq{
\Hgg
&= \min_{M_x \in [-N,N]} \minM \Tr \hr ( \Uh - \ga \Mh)\\
&= \min_{M_x \in [-N,N]} ( \UgM - \ga M_x) \.
}
This means the Legendre transformation\footnote{In the mainstream style of mathematics, $\Hgg$ is {\it minus} the Legendre transformation. Our notation is based on that widely used in thermodynamics~\cite{callen1991thermodynamics}.} of $\UgM$ in terms of $M_x$.
}

\bprop{
$\Hgg$ is a concave function.
}
\bproof{
The Legendre transformation provides a concave function.
}

\bprop{
$\UgM$ is a convex function.
}

\bproof{
For any $\lambda ( 0 \leq \lambda \leq 1), M_- , M_+ (M_- < M_+)$, we fix $\hr_+ \in \argmin_{\hat{\rho}\mid \Tr \hat{\rho} \hat{M}_x^N = M_+} \Uh$ and $\hr_- \in \argmin_{\hat{\rho}\mid \Tr \hat{\rho} \hat{M}_x^N = M_-} \Uh$, which are density matrices minimizing $\Uh$ under the constraint that the $x$-magnetization is $M_\pm$, respectively.
Then, putting $M(\lambda):=(1-\lambda) M_- + \lambda M_+$, we have
\eq{
(1-\lambda) \Ug(M_-) + \lambda \Ug(M_+)
&= \Tr ((1-\lambda) \hr_- +\lambda \hr_+) \Uh \\
&\geq \min_{ \hr\mid \Tr \hr \Mh = M(\lambda)} \Tr \hr \Uh \\
&= \Ug( (1-\lambda) M_- + \lambda M_+) \,
}
which means the convexity of $\UgM$.
}

\bprop{
\label{legendre2}
$\UgM$ is the inverse Legendre transformation of $\Hgg$.
}

\bproof{
It is known that if a function $f$ is the Legendre transformation of a convex function $g$, then the inverse Legendre transformation of $f$ is $g$~\cite{rockafellar1970convex, shimizu2021}.
}

We remark that although the domain of $M_x$ is a finite region $[0,N]$ and that of $\gamma$ is a semi-infinite region $[0, \infty)$, all the aforementioned results are valid for these domains.

\subsection{Self-averaging}
\label{Self-averaging}
In this subsection, we shall show self-averaging of the ground-state energy
\eq{
\frac{\Hgg}{N} \sim \hgg \.
}
with respect to quenched randomness.
Namely, almost all Hamiltonians obtained by random quench have the same ground-state energy density.
Self-averaging allows us to discuss quenched systems only by considering the averaged quantity $\hgg$, not each $\Hgg$.

To this end, we divide the system with Hamiltonian $\Uh$ into $B$ copies of subsystems with equal size $A$ as $N=AB$.
We denote the $k$-th subsystem by $A[k] \kdBA$.
We define the Hamiltonian of $A[k]$ denoted by $\UhAk$ as a restriction of $\Uh$ to subsystem $A[k]$ with removing all the bonds from $A[k]$ to outside $A[k]$. 
We introduce a {\it block-decomposed Hamiltonian} on the same system, which is a product of $\UhAk$ denoted by $\UhAB:=\sumk \UhAk$.
The difference between $\Uh$ and $\UhAB$ is the interaction terms between different subsystems.

The block-decomposed Hamiltonian $\UhAB$ plays a central role in our proof. 
In particular, an argument similar to the law of large numbers is applicable to $\UhAB$, which follows from the fact that $\UhAB$ is a product of i.i.d. random Hamiltonians; $\UhAk \kdB$. 
Since $\UhAB$ is close to $\Uh$, we can derive several self-averaging results in systems with $\Uh$.
Note that this proof idea is a standard technique in the statistical mechanics of random systems (see Ref~\cite{castellani2005spin}).

We first introduce symbols describing an average over quenched randomness and its fluctuation:

\bdfn{
Consider a system with quenched random variables $J$.
Let $X^\NJ$ be a stochastic variable depending on the quenched random variables $J$.
We denote by $[ X^\NJ ]$ the average of $X^\NJ$ with respect to the quenched randomness $J$.
We also denote its root mean square $\sqrt{ [ (X^\NJ)^2 ] }$ by $\SM{X^\NJ}$.
}

Note that $\SM{X^\NJ}$ is a norm (i.e., it satisfies the triangle inequality), which is a direct consequence of Schwarz inequality.

We first bound the root mean square of operator norms of the system Hamiltonian $\Uh$ and the difference between the Hamiltonian and its block-decomposed one; $\Uh - \UhAB$.
The latter quantity can be regarded as surface energy.


\bprop{
\label{three_conditions}
Suppose that $\UhAk \kdB$ are i.i.d. random Hamiltonians of $D$-dimensional systems.
Then, the operator norm of the bulk energy $\Uh$ and the surface energy $\Uh-\UhAB$ are bounded respectively as
\eq{
\SM{ \opnorm{ \Uh } }&\leq \cbulk N \, \\
\SM{\opnorm{\Uh-\UhAB}} &\leq \csurface \AD B \,
}
where $\cbulk$ and $\csurface$ are constants independent of $N$, $A$, and $B$.
}

\bproof{
These bounds are direct consequences of the finite dimensionality of the system.
The bulk energy is bounded as
\eq{
\SM{\opnorm{ \Uh}}
\leq \SM{ \sumi \sumj \abs{J_{ij}}}
\leq \sumi \sumj \SM{J_{ij}}
\leq \cbulk N \,
}
where we used the triangle inequality in the second inequality.
The surface energy is bounded as
\eq{
\SM{\opnorm{\Uh - \UhAB}}
&\leq \SM{ \sumk \sum_{i \in s(A[k])} \sum_{j \notin s(A[k])} \abs{J_{ij}}} \\
&\leq \sumk \sum_{i \in s(A[k])} \sum_{j \notin s(A[k])} \SM{J_{ij}} \\
&\leq \csurface \AD B \,
}
where $s(A[k])$ is a set of sites in subsystem $A[k]$.
}

Now we shall bound the fluctuation of the ground-state energy in terms of quenched randomness.
We first show a slightly weak inequality, and then tighten the inequality by applying the obtained inequality iteratively.

\bprop{
\label{roughHgg}
The standard deviation of $\Hgg$ satisfies the bound
\eq{
\SM{\Hgg - [ \Hgg ]} \leq \cbulk N \. \label{Hgg-weak1}
}
}

\bproof{
For any Hamiltonian $\Uh$, we have the following bound:
\eq{
\abs{ \Hgg + \ga N }
&= \abs{ \minr \Tr[\hr \Hhg] - \minr \Tr[\hr ( - \ga \Mh )] }\\
&\leq \opnorm{ \Hhg + \ga \Mh }\\
&= \opnorm{ \Uh} \.
}
Here, the second line follows from an elementary inequality
\eq{
\label{minop}
\abs{ \minr \Tr \hr \hat{X} - \minr \Tr \hr \hat{Y} } \leq \max_{\hr} \abs{ \Tr \hr \hat{X} - \Tr \hr \hat{Y}} = \opnorm{\hat{X} - \hat{Y}} \.
}
Thus we arrive at the desired inequality:
\eq{
\SM{ \Hgg - [ \Hgg ] } 
&= \SM{ \Hgg + \ga N -[\Hgg + \ga N ]} \\
&\leq \SM{ \Hgg + \ga N}
\leq \SM{ \opnorm{ \Uh } }
\leq \cbulk N \,
}
where the first inequality follows from an elementary fact that $\SM{X-x_0}$ is minimized when $x_0=[X]$, and the last inequality follows from Proposition~\ref{three_conditions}.
}

The above inequality can be tightened by applying the above result to the block-decomposed system $A \# B$ iteratively.

\bprop{
\label{varHgg}
The fluctuation of $\frac{\Hgg}{N}$ vanishes in the thermodynamic limit. 
In particular, for any positive $\ep>0$, we have
\eq{
\SM{ \Hgg - [\Hgg] } = O( N^{1 - 1/ D + \ep}) \.
}
}

\bproof{We start with
\eq{
\SM{ \Hgg - [\Hgg] }
&\leq \SM{ \Hgg - [\HgAB \gb] } \\
&\leq \SM{ \Hgg - \HgAB \gb }+ \SM{ \HgAB \gb - [ \HgAB \gb ] } \. \label{varHgg-mid1}
}
The first inequality follows from that $\SM{X-x_0}$ is minimized when $x_0=[X]$, and the second inequality follows from the triangle inequality.
We first evaluate the first term of the right-hand side of \eqref{varHgg-mid1} as
\eq{
\SM{ \Hgg - \HgAB \gb }
&\leq \SM{ \opnorm{ \Hhg - \HhAB \gb  }} \\
&= \SM{ \opnorm{ \Uh - \UhAB }} \\
&\leq \csurface \AD B \.
}
Here, we used \eqref{minop} in the first inequality, and used Proposition~\ref{three_conditions} in the last inequality.
We next bound the second term of the right-hand side of \eqref{varHgg-mid1}.
Since $E_g^\AkJ \kdB$ are independent random variables, Proposition~\ref{roughHgg} applies to each subsystem, which yields
\eq{
\SM{ \HgAB \gb - [ \HgAB \gb ] }^2
&= \sumk \SM{ E_g^\AkJ \gb - [E_g^\AkJ \gb] }^2 \\
&\leq B \cbulk^2 A^2 \. \label{varHgg-mid2}
}
Combining these two inequalities, we obtain
\eq{
\SM{ \Hgg - [ \Hgg] } \leq \csurface \AD B + \cbulk A B^{1/2} \.
}
Setting $A=N^a$ and $B=N^{1-a}$ with $a = D/(D+2)$, we obtain
\eq{
\SM{ \Hgg - [ \Hgg]}
&\leq (\csurface + \cbulk) N^{1-1/(D+2)} \\
&= O(N^{1-1/(D+2)}) \.\label{Hgg-weak2}
}

We notice that the above inequality \eqref{Hgg-weak2} is stronger than \eqref{Hgg-weak1}.
Therefore, by replacing \eqref{Hgg-weak1} in the derivation of \eqref{varHgg-mid2} by \eqref{Hgg-weak2} (i.e., we use $\SM{ E_g^\AkJ \gb - [E_g^\AkJ \gb] } = O( A^{1-1/(D+2)})$ instead of  $\SM{ E_g^\AkJ \gb - [E_g^\AkJ \gb] } =\cbulk A = O(A)$ in \eqref{varHgg-mid2}), we can obtain a further stronger inequality on $\SM{ \Hgg - [\Hgg] }$. 
By repeating this operation\footnote{Once $\SM{ E_g^\AkJ \gb - [ E_g^\AkJ \gb ]} = O( A^{1-n_m})$ is shown, we can get $\SM{ \Hgg - [ \Hgg ]} = O(A^{1-1/D} B) + O(A^{1-n_m}B^{1/2}) = O(N^{1 - \frac{1}{D+2-2Dn_m}})$ for $a = \frac{D}{D+2-2Dn_m}$. The recurrence formula $n_{m+1}  = \frac{1}{D+2-2Dn_m}$ with the initial term $n_0=0$ has a limit $\lim_{m\to\infty}n_m=1/\max{(D,2)}$.}, 
we finally arrive at
\eq{
\SM{ \Hgg - [ \Hgg] } = O(N^{1 - 1/D + \ep}) \.
}
}

In addition, the existence of the ground-state energy density in the thermodynamic limit can be shown.

\bprop{
\label{limHgg}
The averaged ground-state energy density converges in the thermodynamic limit 
\eq{
\hgg:=\lim_{N \to \infty} \frac{ [ \Hgg ] }{N} \.
}
Moreover, the speed of convergence is evaluated as
\eq{
\frac{[ \Hgg ]}{N} - \hgg = O(N^{-1/D}) \.
}
}
\bproof{
Since $\UhAk \kdB$ are i.i.d. random Hamiltonians, we have $[ \HgAB \gb ] = B [ E_g^{\AoneJ} \gb ]$, which implies 
\eq{
\abs{\frac{[\Hgg]}{N} - \frac{[E_g^{\AoneJ} \gb]}{A}} = \abs{\frac{[ \Hgg ]}{N} - \frac{[\HgAB \gb]}{N} } \leq \csurface A^{-1/D} \.
}
This shows that $a_N:=\frac{ [ \Hgg ] }{N}$ is a Cauchy sequence and hence converges.
}

We finally prove the self-averaging of the ground-state energy.

\begin{thm}[Self-averaging of $\Hgg$]
\label{pointwise_hgg}
For any $\ga$, the ground-state energy density $\frac{\Hgg}{N}$ converges to $\hgg$ in mean square:
\eq{
\SM{ \frac{\Hgg}{N} - \hgg } = O(N^{ - 1/D + \ep}) \.
}
\end{thm}


\bproof{Combining Proposition~\ref{varHgg} with Proposition~\ref{limHgg}, we easily have 
\eq{
\SM{ \Hgg - N\hgg) }
&\leq \SM{ \Hgg - [ \Hgg ] } + \SM{ [ \Hgg ] - N\hgg } \\
&\leq \SM{ \Hgg - [ \Hgg ] } + \csurface N^{1-1/D} \\
&= O(N^{1-1/D+\ep}) \,
}
which is equivalent to the desired result.
}

\subsection{Uniform self-averaging}
\label{uniform-self-averaging}
In this subsection, we will show uniform self-averaging (i.e., self-averaging as a function of $\gamma$), which is a stronger condition than the self-averaging discussed in the previous subsection. The key idea for the proof of uniform self-averaging is to put many \textit{regularity checkpoints} on the $\ga$-axis. To demonstrate self-averaging for any $\ga$, we employ self-averaging at the nearest regularity checkpoint of $\gamma$ and evaluate the speed of convergence at $\gamma$.
Since it is not easy to show uniform self-averaging in the half-infinite region $[0, \infty)$ directly, we set the domain of $\gamma$ in $\Hgg$ as $[0, \gbi]$ for $\gbi$ that diverges slowly as $N$ increases.


We start by showing that $\Hgg$ is Lipschitz continuous.

\bprop{
\label{gadiff}
For any $\ga_1 ,\ga_2$ and any instance, the difference between $\Hgg$ (and $\hgg$) with $\ga_1$ and $\ga_2$ is bounded as
\eq{
\abs{ \Hg(\ga_1) - \Hg(\ga_2) } &\leq N \abs{ \ga_1 - \ga_2 } \, \label{gadiff-1} \\
\abs{ e_g(\ga_1) - e_g(\ga_2)} &\leq  \abs{\ga_1 - \ga_2} \.
}
}
\bproof{

The first inequality of \eqref{gadiff-1} follows from \eqref{minop} as $\abs{ \Hg(\ga_1) - \Hg(\ga_2) } \leq \opnorm{ \Hh(\ga_1) - \Hh(\ga_2) } = N \abs{\ga_1 - \ga_2}$. The same argument holds for their mean and under the thermodynamic limit.}

\begin{thm}[Uniform self-averaging of $\Hgg$]
\label{uniform_hgg}
The ground-state energy density $\frac{\Hgg}{N}$ converges uniformly on $[0,\gbi]$ in mean square:
\eq{
\SM{ \maxgl \abs{ \frac{\Hgg}{N} - \hgg } }
= O(N^{-2/(5D)+\ep}) \,
}
where we set $\gbi = N^{1/(5D)}$.
\end{thm}

\bproof{
For convenience, we suppose that $N^{1/D+1/(5D)}$ is an integer.
Corresponding to integers $w = 1, \dots , N^{1/D+1/(5D)}$, we define the regularity checkpoints and their covering intervals as
\eq{
\ga_w &= (w - 1/2) N^{-1/D}, \\
I_w &= \[(w-1) N^{-1/D}, w N^{1/D}\] \.
}
With noting that $\abs{\ga - \ga_w} \leq \frac{N^{-1/D}}{2}$ for any $\gamma\in I_w$, we have
\eq{
&\abs{ \Hgg - N \hgg } \\
\leq& \abs{ \Hgg - \Hg(\ga_w) } + \abs{ \Hg(\ga_w) - N e_g(\ga_w)}
+ \abs{ Ne_g(\ga_w) - Ne_g(\ga)} \\
\leq& \abs{ \Hg(\ga_w) - N e_g(\ga_w)} + N^{1-1/D} \,
}
where we used Proposition~\ref{gadiff} in the second inequality.
Hence the maximum deviation of ground-state energy is bounded as
\eq{
\SM{\max_{\ga \in I_w} \abs{ \Hgg - N\hgg }}
\leq& \SM{ \Hg(\ga_w) - Ne_g(\ga_w) } + N^{1-1/D} \. \label{uniform_hgg-mid1}
}
Using this relation, we arrive at the desired result:
\eq{
\SM{ \max_{\ga \in [0,\gbi]} \abs{ \Hgg - N\hgg } }^2
&= \SM{ \max_{w=1}^{N^{1/D+1/(5D)}} \max_{\ga \in I_w} \abs{ \Hgg - N\hgg } }^2 \\
&\leq \sum_{w=1}^{N^{1/D+1/(5D)}} \SM{ \max_{\ga \in I_w} \abs{ \Hgg - N\hgg } }^2 \\
&\leq \sum_{w=1}^{N^{1/D+1/(5D)}} \( \SM{ \Hg(\ga_w) - Ne_g(\ga_w) } + N^{1-1/D} \)^2 \\
&= O( N^{2-1/D+1/(5D)+\ep}) \.
}
In the first inequality we used the following simple relation for nonnegative $L_w$;
\eq{
\SM{\max_w L_w}^2 = \[ \max_w L_w^2 \] \leq \[ \sum_w L_w^2 \] = \sum_w \SM{L_w}^2 \,
\label{Lw}
}
in the second inequality we used \eqref{uniform_hgg-mid1}, and in the last inequality we used Proposition~\ref{pointwise_hgg}.
}

We proceed to the uniform self-averaging of $\Ugm$.
To prove this, we introduce the inverse Legendre transformation of $\hgg$, to which the ground-state energy density $\frac{\Ugm}{N}$ converges.

\bdfn{The inverse Legendre transformation of $\hgg$ is defined as
\eq{
\ugm &= \sup_{\ga \in (-\infty, \infty)} ( \hgg + \ga m_x) \.
}
}

\begin{thm}[Uniform self-averaging of $\Ugm$]
\label{uniform_ugm}
$\frac{\Ugm}{N}$ converges
uniformly on $\[0,\mbi \]$ in mean square:
\eq{
\SM{ \maxml \abs{ \Ugm - N\ugm } }
= O(N^{1-2/(5D)+\ep}) \,
}
where $\mbi = 1 - \frac{2}{N^{1+1/(5D)}} \opnorm{\Uh}$.
\end{thm}

\bproof{
Consider a pair $(m_x, \ga)$ satisfying 
\eq{
\Ugm = \Hgg + \ga N m_x \. \label{mx-gamma-rel}
}
Since $\Hgg$ is the Legendre transformation of $\Ugm$ and its minimum is achieved at $M_x=Nm_x$, we have
\eq{
\Ugm - \ga Nm_x &\leq \Ug(N) - \ga N \,
}
and thus
\eq{
\ga ( 1 - m_x ) &\leq \frac{\Ug(N) - \Ugm}{N} \leq \frac{2\opnorm{\Uh}}{N}
}
holds. 
It follows that $m_x \leq \mbi$, then the corresponding $\gamma$ with \eqref{mx-gamma-rel} satisfies $\ga \leq \gbi=N^{1/5D}$.
Hence, the domain of $\gamma$ in the maximization in the Legendre transformation of $\Hgg$ and $\hgg$ can be narrowed to $[0,\gbi]$:
\eq{
\Ugm =& \maxgl ( \Hgg + \ga N m_x) \, \\
\ugm =& \maxgl ( \hgg + \ga m_x) \.
}
Then, the difference between energy of a single instance and its average after taking the thermodynamic limit is evaluated as
\eq{
&\abs{ \Ugm - N\ugm } \\
=& \abs{ \maxgl ( \Hgg + \ga N m_x) - N\ugm } \\
\leq& \abs{ \maxgl ( N \hgg + \ga N m_x) - N \ugm }+ \abs{\maxgl ( \Hgg - N \hgg )} \\
\leq& \maxgl \abs{ \Hgg - N \hgg } \.
}
Plugging Proposition~\ref{uniform_hgg} into the above inequality, we have the desired result:
\eq{
\SM{ \maxml \abs{ \Ugm - N \ugm } }
&\leq
\SM{ \maxgl \abs{ \Hgg - N \hgg }} \\ &= O(N^{1-2/(5D) + \ep}) \.
}
}

\subsection{Antiferromagnetic fluctuations}
\label{subaff}

Suppose that the $x$-magnetization $M_x$ shows a first-order phase transition (i.e., discontinuous jump) at some $\gamma$.
At this point $\hgg$ is no longer differentiable, and $\ugm$ is convex but not strictly convex.
Our idea to avoid the first-order phase transition in $M_x$, based on the above observation, is adding a strictly convex function to $\ugm$.
By construction, the modified $\ugm$ is strictly convex, and $\hgg$ has no singularity.

In particular, we add a quantum antiferromagnetic fluctuations term $\alN (\Mh)^2$ to the Hamiltonian, which we denote by $\Hhga$.
Correspondingly, we denote the minimum energy conditioned by $M_x$ by $\UgMa$.
Then, Theorem~\ref{uniform_ugm} suggests
\eq{
\frac{ \Ugma }{N} \sim \ugma := \ugm + \al m_x^2 \.
}
Since $\ugma$ is a strictly convex function, quantum first-order phase transition in $M_x$ does not occur.

A nontrivial step in the aforementioned proof outline is connecting $\ev{ (\Mh)^2 }$ and $\ev{\Mh}^2$, since these two are in general not equal; $\ev{ (\Mh)^2 } \neq \ev{\Mh}^2$. 
In fact, these two are inequivalent in some long-range interacting systems (e.g., $p$-spin model with large $p$, discussed in~\cite{seki2012quantum}).
On the other hand, we can prove $\ev{ (\Mh)^2 }\simeq \ev{\Mh}^2$ in short-range interacting systems.
This is the main task in this subsection.
We note that our argument does not hold for $\al < 0$.

\bdfn{
\label{qaaff}
For $\al \in ( 0, \infty) $, we introduce Hamiltonians and related quantities corresponding to QA-AFF denoted by
\eq{
\Hhga &= \Hhg + \alN ( \Mh )^2 \, \\
\Uha &= \Uh + \alN ( \Mh )^2 \, \\
\UgMa &= \min_{\hr \mid \Tr \hr \Mh = M_x}  \Tr \hr \Uha \, \\
\ugma &= \ugm + \al m_x^2 \.
}
}

We first prove the uniform self-averaging of $\Ugma$.

\bprop{
The ground-state energy density $\frac{\Ugma}{N}$ converges uniformly to $\ugma$ on $[0,\mbi]$ in mean square:
\eq{
\SM{ \max_{m_x \in [ 0 , \mbi]}
\abs{ \frac{\Ugma}{N} - \ugma } }
= O(N^{-2/(5D)+\ep}) \.
}
}
\bproof{
We first derive the following bound:
\eq{ 
\abs{ \UgAB \mxa - \UgAB \mx - \al N m_x^2} \leq \al A \. \label{U-a-eval}
}
The elementary inequality $\Tr \hr (\Mh)^2 - (\Tr \hr \Mh)^2 \geq 0$ implies
\eq{
\label{AsharpB-1}
\UgAB \mxa \geq \UgAB \mx + \al N m_x^2 \.
} 
Fix $\hr' \in \argminm \Tr \hr \UhAB$, which has $x$-magnetization $Nm_x$ and minimizes the energy $\UhAB$.
We construct a state from $\hr'$ by removing all the correlation between subsystems:
\eq{
\hr_k &= \Tr_{i \notin s(A[k])} \hr' \, \\
\hr_\otimes &= \bigotimes_{k=1}^B \hr_k \.
}
By construction, $\hr_\otimes$ is separable into subsystems, and $\hr_\otimes$ also minimizes $\UhAB$ with $x$-magnetization equal to $Nm_x$: $\hr_\otimes \in \argminm \Tr \hr \UhAB$. 
The AFF term in $\hr_\otimes$ can be directly estimated as follows:
\eq{
&\Tr \hr_\otimes ( \Mh )^2 \\
=& \sumk \suml \Tr \hr_\otimes \MhAk \MhAl \\
=& \sumk \suml \Tr \hr_\otimes \MhAk \Tr \hr_\otimes \MhAl + \sumk \left( \Tr \[ \hr_\otimes ( \MhAk )^2 \] - \left( \Tr \hr_\otimes (\MhAk)\right)^2 \right) \\
\leq& N^2 m_x^2 + B A^2 \,
}
which implies a relation evaluating the difference between $\UgAB$ with and without the AFF term:
\eq{
\label{AsharpB-2}
\UgAB \mx
&= \Tr \hr_\otimes \UhAB \\
&= \Tr \hr_\otimes \UhAB(Nm_x;\al) - \alN \Tr \[\hr_\otimes ( \Mh )^2\] \\
&\geq \UgAB \mxa - \al N m_x^2 - \al A \.
}
Using the two inequalities \eqref{AsharpB-1} and \eqref{AsharpB-2}, we arrive at \eqref{U-a-eval}.

Now we fix $A = N^{a'}$ with $a' = 1/(D+1)$ in \eqref{U-a-eval}.
Combining \eqref{U-a-eval}, Proposition~\ref{three_conditions}, and Proposition~\ref{uniform_ugm} with recalling $N\ugma=N\ugm +\alpha Nm_x^2$, we obtain the desired result:
\eq{
&\SM{ \maxml \abs{ \Ugma - N\ugma }} \\
\leq&
\SM{ \maxml \abs{ \Ugma - \UgAB \mxa }} \\
&+ \SM{ \maxml \abs{ \UgAB \mxa - \UgAB \mx - \alpha N m_x^2 }} \\
&+
\SM{ \maxml \abs{ \UgAB \mx - \Ugm }} \\
&+ \SM{ \maxml \abs{ \Ugm - N\ugm }} \\
\leq& \SM{ \maxml \abs{ \Ugm - N\ugm }} + \al A + 2 \csurface \AD B \\
=& O( N^{ 1 - 2/(5D) + \ep} ) \.
}
Here, the first inequality follows from the triangle inequality.
}

\bdfn{
We denote by $m_*(\ga;\al)$ the unique argument that minimizes $\ugma - \ga m_x$.
We also define $M_*^\NJ(\ga; \al)$ as the expectation value of $\Mh$ in a ground state of $\Hhga$ (i.e., $\ev{\Mh}{g}$, where $\ket{g}$ is a ground state of $\Hhga$).
For brevity, we sometimes drop the arguments $\ga$ and $\al$ in $m_*(\ga;\al)$ and $M_*^\NJ(\ga; \al)$, and simply write $m_*$ and $M_*^\NJ$.
}


Since $\ugma$ is a strictly convex function, $m_*(\ga; \al)$ is a continuous function\footnote{In fact, $m_*(\ga; \al)$ is Lipschitz continuous with constant $1/(2\al)$, which means that there is no quantum second-order transition in transverse magnetization either.} with respect to $\ga$ for any $\al>0$. 
We shall show that $\frac{M_*^{N:J}}{N}$ converges to a continuous function $m_*(\ga; \al)$, which completes the proof of our main result.

\begin{thm}[Absence of the QFOT]
\label{uniform_M}
$\frac{M_*^{N:J}}{N}$ (as a function of $\ga$) converges uniformly on $[0,\infty)$ in mean square:
\eq{
\SM{ \sup_{\ga \in [0,\infty)} \max_{ \ket{g} \in G(\Hhga)} \abs{ \frac{\ev{\Mh}{g}}{N} - m_*(\ga; \al) } } = O(N^{-1/(5D) + \ep}) \. \label{main-ineq}
}
\end{thm}

\bproof{
Note that 
\eq{\SM{1-\mbi} = \frac{\SM{2\opnorm{\Uh}}}{N^{1+1/(5D)}} \leq \frac{2\cbulk}{N^{1/(5D)}}\. \label{mbi-bound}} 
We decompose the domain of $M_x$, $[0,N]$, into two regions, $I_1:=[0 , N\mbi]$ and $I_2:=[N\mbi, N]$.
Promising $\sup \emptyset = 0$ for convenience, we evaluate the square of the left-hand side of \eqref{main-ineq} (multiplied by $N$) as
\eq{
&\SM{ \sup_{\ga \in [0,\infty)} \max_{ \ket{g} \in G(\Hhga)} \abs{ \ev{\Mh}{g} - Nm_*(\ga; \al) } }^2 \\
\leq& \SM{ \supll \abs{  M_*^\NJ - Nm_* }}^2 + \SM{ \suplr \abs{  M_*^\NJ - Nm_* }}^2 \\
&+ \SM{ \suprl \abs{  M_*^\NJ - Nm_* }}^2 + \SM{ \suprr \abs{  M_*^\NJ - Nm_* }}^2 \.
\label{last-max}
}
Here, we used \eqref{Lw}. We shall evaluate these four terms.

Before going to the evaluation, we introduce a useful relation that if functions $P,Q,R,S,T$ satisfy $P \leq Q$ and $S + T^2\leq R$, then
\eq{
\SM{ \max \abs{T}}^2 \leq \SM{ \max \abs{ P-R } } + \SM{ \max \abs{ Q-S } }  \label{last-lemma}
}
is satisfied.
This relation is easily confirmed as
\eq{
\SM{ \max \abs{T}}^2 &\leq [ \max \abs{ R - S } ] \\
&\leq [ \max ( \abs{P-R} + \abs{Q-S} )] \\
&\leq [ \max \abs{ P-R } ] + [ \max \abs{ Q-S } ] \\
&\leq \SM{ \max \abs{ P-R } } + \SM{ \max \abs{ Q-S } } \. 
}

Now we evaluate the four terms in \eqref{last-max}.
To evaluate the first term of \eqref{last-max}, we consider
\eq{
P &= \Ug(M_*^\NJ ; \al) - \ga M_*^\NJ \, \\
Q &= \Ug(Nm_* ; \al) - \ga N m_* \, \\
R &= Nu_g\( \frac{M_*^\NJ}{N}; \al\) - \ga M_*^\NJ \, \\
S &= Nu_g(m_*; \al) - \ga N m_* \, \\
T &= \( \alN \)^{1/2} \abs{ M_*^\NJ - Nm_* } \.
}
Here, $P \leq Q$ follows from the fact that $M_*^\NJ$ is the minimizer of $\UgMa - \ga M_x$. Also, $S + T^2\leq R$ follows from the fact that $m_*$ is the minimizer of $\ugma - \ga m_x$ and that the second derivative of $\ugma$ is at least $2 \al$. Thus, it follows from \eqref{last-lemma} that
\eq{
&\SM{ \supll \abs{  M_*^\NJ - Nm_* }}^2 \\
\leq& \frac{N}{\al} \SM{ \supll \abs{ \Ug(M_*^\NJ; \al) - Nu_g\( \frac{M_*^\NJ}{N} ; \al\) } } \\
&+ \frac{N}{\al} \SM{ \supll \abs{ \Ug(Nm_*; \al) - Nu_g(m_* ; \al) } } \\
\leq& \frac{N}{\al} \SM{ \sup_{\ga, \ket{g} \mid M_*^\NJ \in I_1} \abs{ \Ug(M_*^\NJ; \al) - Nu_g\( \frac{M_*^\NJ}{N} ; \al\) } } \\
&+ \frac{N}{\al} \SM{ \sup_{\ga, \ket{g} \mid Nm_* \in I_1} \abs{ \Ug(Nm_*; \al) - Nu_g(m_* ; \al) } } \\
=& O(N ^ {2 - 2/(5D) + \ep}) \.
}
In the last line, we used  Proposition~\ref{uniform_ugm}.

To evaluate the second term of \eqref{last-max}, we use \eqref{last-lemma} with
\eq{
P &= \Ug(N\mbi ; \al) - \ga N\mbi \, \\
Q &= \Ug(Nm_* ; \al) - \ga N m_* \, \\
R &= Nu_g(\mbi; \al) - \ga N\mbi \, \\
S &= Nu_g(m_*; \al) - \ga N m_* \, \\
T &= \( \alN \)^{1/2} \abs{ N \mbi - Nm_* } \.
}
Here, $P\leq Q$ follows from (i) the fact that $M_*^\NJ$ is the minimizer of $\UgMa - \ga M_x$, (ii) the convexity of $\UgMa - \ga M_x$, and (iii) an assumption that $Nm_* \leq N\mbi \leq M_*$. Also, $S + T^2 \leq R$ holds by the same argument as in the first term. Thus, it follows from \eqref{last-lemma} that
\eq{
&\SM{ \suplr \abs{  N\mbi - Nm_* }}^2 \\
\leq& \frac{N}{\al} \SM{ \abs{\Ug(N\mbi;\al) - N u_g(\mbi;\al)}} \\
&+ \frac{N}{\al} \SM{ \sup_{\ga,\ket{g} \mid Nm_* \in I_1} \abs{\Ug(Nm_*;\al) - Nu_g(m_*;\al)}} \\
=& O(N ^ {2 - 2/(5D) + \ep}) \.
}
In the last line, we used Proposition~\ref{uniform_hgg} and the fact that $N\mbi \in I_1$. Using this inequality and \eqref{mbi-bound}, we have
\eq{
&\SM{ \suplr \abs{  M_*^\NJ - Nm_* }} \\
\leq& \SM{ \suplr \abs{  N\mbi-Nm_* }} + \SM{N(1-\mbi)} \\
=& O(N^{1-1/(5D)+\ep}) \.
}

To evaluate the third term of \eqref{last-max}, we use \eqref{last-lemma} with
\eq{
P &= \Ug(M_*^\NJ ; \al) - \ga M_*^\NJ \, \\
Q &= \Ug(N\mbi ; \al) - \ga N \mbi \, \\
R &= Nu_g(M_*^\NJ/N; \al) - \ga M_*^\NJ \, \\
S &= Nu_g(\mbi; \al) - \ga N \mbi \, \\
T &= \( \alN \)^{1/2} \abs{ M_*^\NJ - N\mbi } \.
}
Here, $P \leq Q$ holds by the same argument as in the first term. Also, $S + T^2 \leq R$ follows from (i) the fact that $m_*$ is the minimizer of $\ugma - \ga m_x$, (ii) the fact that the second derivative of $\ugma$ is at least $2 \al$ and (iii) an assumption that $M_* \leq N\mbi \leq Nm_*$. Thus, it follows from \eqref{last-lemma} that
\eq{
&\SM{ \suprl \abs{  M_*^\NJ - N\mbi }}^2 \\
\leq& \frac{N}{\al} \SM{ \sup_{\ga,\ket{g} \mid M_*^\NJ \in I_1} \abs{\Ug(M_*^\NJ;\al) - N u_g \(\frac{M_*^\NJ}{N};\al \)}} \\
&+ \frac{N}{\al} \SM{ \abs{\Ug(N\mbi;\al) - Nu_g(\mbi;\al)}} \\
=& O(N ^ {2 - 2/(5D) + \ep}) \.
}
Consequently, we have
\eq{
&\SM{ \suprl \abs{  M_*^\NJ - Nm_* }} \\
\leq& \SM{ \suprl \abs{ M_*^\NJ - N\mbi }} +\SM{N(1-\mbi)} \\
=& O(N^{1-1/(5D)+\ep}) \.
}

The last term of \eqref{last-max} is simply bounded as
\eq{
\SM{ \suprr \abs{  M_*^\NJ - Nm_* }} 
&\leq \SM{N ( 1 - \mbi)} = O(N^{1-1/(5D)}) \.
}

Combining these four inequalities, we complete the proof of Theorem~\ref{main_theorem}.
}

\section{Discussion}
\label{discussion}

We have proved that the quantum annealing (QA) for finite-dimensional spin-glass systems does not show a quantum first-order transition in transverse magnetization (QFOT) by adding the antiferromagnetic fluctuations (AFF) term.
This result holds for any spin-glass system as long as the system is in finite dimensions and its quenched randomness is sampled from a shift-invariant probability distribution.
For simplicity of explanation, we assume that the interaction in $\Uh$ is two-body and the boundary is an open boundary condition, but our result applies to more general systems.
In fact, our proof relies only on Proposition~\ref{three_conditions} (finite dimensionality) and the fact that subsystems are i.i.d.
Thus, if these two conditions are satisfied, our result also applies to systems with the periodic and closed boundary conditions as well as those with local fields and short-range $p$-body interactions.

\paragraph{Key ideas in our proof}

We first elucidate the power of uniform self-averaging.
Uniform self-averaging is an important concept for discussing the absence of phase transitions. 
Applying an argument analogous to Chebyshev's inequality, we show that the function $\frac{M_*^\NJ(\ga)}{N}$ is in the $\sup$-norm neighborhood of the function $m_*(\ga)$ for almost all $J$. Conventional self-averaging alone, which corresponds to pointwise convergence, cannot eliminate the possibility that there is a discontinuous jump in each instance with different transition points depending on instances.
On the other hand, uniform self-averaging indeed prohibits this unwanted possibility.

Next, we discuss the role of the AFF term.
Thanks to the description with $\UgM$, our approach makes the meaning of the AFF term ($(\Mh)^2$ term) much clearer than in the original paper of QA-AFF~\cite{seki2012quantum}. 
Namely, the AFF term strengthens the convexity in $\ugm$ and ensures that it is strictly convex, not merely convex, and this fact shows that the transverse magnetization is continuous with respect to $\ga$.
Similar arguments can be seen in some papers in statistical mechanics~\cite{hetherington1987solid, yoneta2019squeezed}, where the difficulties associated with first-order phase transitions are solved by devising the shape of the ensemble. 
However, we should notice the discrepancy $\ev{ (\Mh)^2 } \neq \ev{\Mh}^2$, which prevents a direct analogous argument. 
In particular, the procedure to obtain a narrowly convex function $\ugma$ as presented in Section~\ref{subaff} does not always work well for long-range interacting systems, e.g., $p$-spin model with large $p$~\cite{seki2012quantum}.

\paragraph{Implications to the hardness of QA}

It is numerically well known that the QA for hard combinatorial optimization problems fails at some point in the QA process.
As explained in the Introduction, the role of the QFOT in the failure of the QA is controversial.
Our result says that the QFOT in QA for finite-dimensional spin glasses can be removed by adding the AFF term.
We expect that the QFOT in {\it any} extensive sum of local observables $\hat{A}$ is avoidable by a slight modification of QA.
If the observable $\hat{A}$ does not contain $z$-magnetization $\hat{\sigma}^z$, a slight extension of our argument leads to the desired consequence by simply adding the fluctuations term $\frac{\hat{A}^2}{N}$.
On the other hand, if $\hat{A}$ contains $z$-magnetization, our estimation $\langle \hat{A}^2\rangle=O(1)$ for the optimal solution no longer holds, and some additional ideas are necessary, which is left for future research. 


We emphasize that our result does not claim that the QA in finite-dimensional spin glasses succeeds and that the corresponding ground-state search problem can be efficiently solved. 
A more plausible scenario suggested by our result is that the QA in finite-dimensional spin glasses fails for different reasons from the QFOT.
One candidate is the {\it glassy bottlenecks}, which are undetectable by the usual macroscopic observables.
It is shown in Ref~\cite{knysh2016zero} that some models with a transverse magnetic field have exponentially small gaps in the glass phase rather than at the phase transition point.
 The arrangement of the ground state from one glassy state to another glassy state can make QA less efficient.
Our result supports this picture.



However, it should be clarified that two statements can be reconciled: (i) the ground-state search problem for finite-dimensional spin glasses is efficiently solvable by QA-AFF, and (ii) no quantum computer can solve NP-hard problems efficiently.
This apparent contradiction is resolved for the following reason:
Statement (i) concerns the average-case hardness, which means that almost all instances of spin glasses can be solved efficiently.
In contrast, statement (ii) concerns the worst-case hardness, claiming that for any quantum computer, there exists at least one instance that cannot be solved efficiently.
Hence, it is possible that the ground state search problem for finite-dimensional spin glasses, which is an $\NPh$ problem, is typically easy and rarely has hard instances.

\paragraph{Future works}
Note that it is difficult to simulate a QA-AFF classically because the AFF term is non-stoquastic~\cite{albash2019role} and gives rise to a negative sign problem. 
It is an open question whether we can avoid quantum first-order phase transition only by adding stoquastic terms to a QA.

Also, the exact minimization problem for finite-dimensional spin glass for $D \geq 3$ is $\NPh$, though it is easy to solve the minimization problem that allows any errors proportional to the system size $N$. On the other hand, the PCP theorem~\cite{arora1998proof, dinur2007pcp, haastad2001some} implies that there are problems for which even approximate minimization is $\NPh$.
It is an open question whether there exist QA-AFF or other extensions of QA for such problems where no quantum first-order phase transitions occur.

\bmhead{Acknowledgements}
We thank \mbox{Masayuki Ohzeki}, \mbox{Jun Takahashi}, \mbox{Yasushi Yoneta}, \mbox{Yuuya Chiba} and \mbox{Hidetoshi Nishimori} for valuable discussions.
NS is supported by JSPS KAKENHI Grants-in-Aid for Early-Career Scientists Grant Number JP19K14615. 
KH is supported by JST Grant Number JPMJPF2221 and JSPS KAKENHI Grant Number 23H01095. 

\bmhead{Competing interests}
The authors declare no competing interest.

\bmhead{Data availability}
No datasets were generated or analyzed during the current study.


\bibliography{qa-bibliography}


\end{document}